\documentclass[sigconf]{acmart}

\AtBeginDocument{%
  }

\copyrightyear{2026}
\acmYear{2026}
\setcopyright{cc}
\setcctype{by}
\acmConference[IDE '26]{3rd International Workshop on Integrated Development Environments }{April 12--18, 2026}{Rio de Janeiro, Brazil}
\acmBooktitle{3rd International Workshop on Integrated Development Environments (IDE '26), April 12--18, 2026, Rio de Janeiro, Brazil}
\acmDOI{10.1145/3786151.3788608}
\acmISBN{979-8-4007-2384-1/2026/04}

\usepackage{cleveref}
\usepackage{float}
\usepackage{xcolor}
\usepackage{subcaption}

\begin{document}

\title{Control Models for In-IDE Code Completion}
\subtitle{Saving Inference Costs While Improving Completion Quality Metrics}

\author{Aral de Moor}
\orcid{0009-0003-5105-0518}
\authornote{Authors contributed equally to this research.}
\affiliation{%
  \institution{JetBrains}
  \city{Amsterdam}
  \country{The Netherlands}
}
\email{aral.de.moor@jetbrains.com}

\author{Yana Hrynevich}
\orcid{0009-0004-5325-5565}
\authornotemark[1]
\affiliation{%
  \institution{JetBrains}
  \city{Prague}
  \country{Czech Republic}
}
\email{yana.hrynevich@jetbrains.com}

\author{Hleb Badzeika}
\orcid{0009-0003-8053-5135}
\authornotemark[1]
\affiliation{%
  \institution{JetBrains}
  \city{Warsaw}
  \country{Poland}
}
\email{hleb.badzeika@jetbrains.com}

\author{Vladyslav Furda}
\orcid{0009-0003-4722-1169}
\authornotemark[1]
\affiliation{%
  \institution{JetBrains}
  \city{Prague}
  \country{Czech Republic}
}
\email{vladyslav.furda@jetbrains.com}

\author{Marko Kojic}
\orcid{0009-0000-1348-5930}
\authornotemark[1]
\affiliation{%
  \institution{JetBrains}
  \city{Belgrade}
  \country{Serbia}
}
\email{marko.kojic@jetbrains.com}

\author{Artem Savelev}
\orcid{0009-0002-5452-1082}
\authornotemark[1]
\affiliation{%
  \institution{JetBrains}
  \city{Munich}
  \country{Germany}
}
\email{artem.savelev@jetbrains.com}

\author{Kostadin Cvejoski}
\orcid{0009-0003-6976-3997}
\authornotemark[1]
\affiliation{%
  \institution{JetBrains Research}
  \city{Bonn}
  \country{Germany}
}
\email{kostadin.cvejoski@jetbrains.com}

\author{Darya Rovdo}
\orcid{0009-0001-9073-3296}
\authornotemark[1]
\affiliation{%
  \institution{JetBrains}
  \city{Amsterdam}
  \country{The Netherlands}
}
\email{darya.rovdo@jetbrains.com}

\author{Ekaterina Garanina}
\orcid{0009-0005-8042-5651}
\authornotemark[1]
\affiliation{%
  \institution{JetBrains}
  \city{Yerevan}
  \country{Armenia}
}
\email{ekaterina.garanina@jetbrains.com}

\renewcommand{\shortauthors}{de Moor et al.}

\begin{abstract}
  We introduce control models for LLM-powered code completion in JetBrains IDEs: ML classifiers which trigger inference and filter the generated suggestions to better align them with users and reduce unnecessary requests.
  To this end, we evaluate boosting- and transformer-based architectures on an offline dataset of real code completions with $n=98$ users.
  We further evaluate the offline classification performance of our boosting-based approach on a range of syntactically diverse languages; and perform an A/B study in a production environment where they 
  improve completion efficiency and quality metrics.
  With this study, we hope to demonstrate the potential in using auxiliary models for smarter in-IDE integration of LLM-driven features, highlight fruitful future directions, and open problems.

\end{abstract}

\ccsdesc[500]{Human-centered computing~Human computer interaction (HCI)}
\ccsdesc[500]{Computing methodologies~Supervised learning by classification}

\keywords{IDE, Machine Learning, Classification, Usability, Interaction, Code Completion, Large Language Model}


\received{27 October 2025}
\received[accepted]{18 December 2025}
\received[revised]{26 January 2026}

\maketitle

\begin{figure}[H]
  \centering
  \includegraphics[width=1\linewidth]{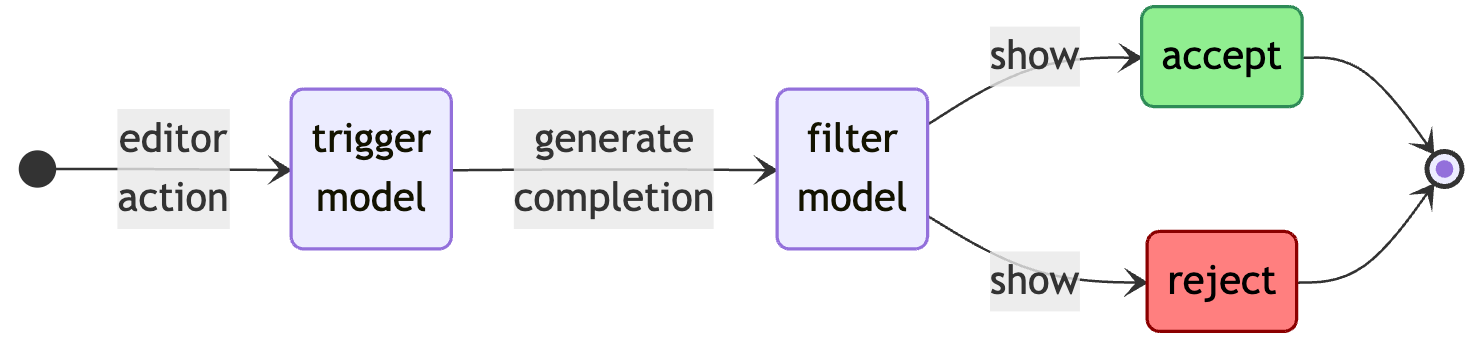}
  \caption{Control Models Problem Setting: Completion Flow with Positive (Accept) and Negative (Reject) Targets.}
  \label{fig:control-models-flow}
  \Description[Control Models Problem Setting]{
    After an editor action, the trigger model is invoked. 
    If the trigger model passes the event, a completion is generated, which is then passed to the filter model. 
    Only when the filter model passes the completion, is it shown to the user. 
    This provides our labels for the dataset, an event is positive if it is accepted by the user, and negative otherwise.
  }
\end{figure}

\section{Introduction}
When it comes to code completion powered by Large Language Models (LLMs),
suggestions with incorrect code, poor context-awareness, or bad timing can be distracting to the developer~\cite{RovdoKeynote}.
Furthermore, misalignment with their 
state of mind~\cite{Barke2022} 
or experience level~\cite{Prather2023} 
can be disruptive to their flow.
At best, this results in wasteful LLM inference
~\cite{Ziegler2022, Sun2025, MozannarWhen}, 
but often leads to \textit{counterproductive} interactions 
~\cite{Vaithilingam2022, Prather2023, Barke2022, Sergeyuk2025}.
As LLMs are becoming more integrated into developer workflows, it is necessary to think beyond aligning content itself – and also consider how users interact with it.

We introduce the term \textit{Control Models}, which 
\textit{modify human-AI interactions with the goal of better aligning LLMs with developers' flow}.
To this end, control models leverage explicit in-IDE information that is not available to the language model,
namely in-IDE telemetry such as previous actions, typing speed and caret scope information, as well as the outputs of static analyses
(e.g.~\cite{MozannarWhen, deMoor2024, Cipollone2025, RovdoBlog}).
We intentionally settle on a definition that does not mention a specific IDE feature, as we believe that the advancements in LLM use-cases come hand-in-hand with an increased potential to disrupt user flows; and expect control models to be generally relevant.

In the context of this paper though, we consider the application of control models to inline ``grey-text'' completion
~\cite{Pavlichenko2025,SemenkinFLCC}.
This involves limiting the ``suggestion space'' of inline completions to align better with developer workflows, using, specifically, smarter \textit{triggers} for generation
~\cite{deMoor2024, Sun2025, MozannarWhen},
and smarter \textit{filters} thereafter
~\cite{MozannarWhen}.
This is motivated by only 31\% of inferences materialising into shown completions, 
and only 31\% of those shown completions being accepted, 
as well as by user reports of 
distraction~\cite{Prather2023, RovdoKeynote} 
and unnecessary context-switches~\cite{Barke2022, Vaithilingam2022}.
See \Cref{fig:control-models-flow} for a diagram of the problem setting.


To summarise, our contributions are:
\begin{itemize}
  \item An offline analysis and online AB study of lightweight, boosting-based trigger and filter classifier models, demonstrating that they can save around 20\% of inference requests while improving completion quality metrics in JetBrains IDEs.
  \item Large-scale offline empirical evidence highlighting the potential of transformer models in this area, which successfully combine contextual code understanding with additional in-IDE feature modalities; but also the obstacles regarding latency, privacy, and integration we faced.
  \item A discussion underscoring the most fruitful directions for future work on similar control models. This includes our development of evaluation metrics towards ones that measure the proclaimed productivity gains over a longer term, an exploration of alternative model architectures, and observations that highlight the need to design for user-personalisation.
\end{itemize}

\section{Related Work}

\subsection{Productive Human-AI Interaction}
AI Assistance is commonplace in IDEs, powering widely used features such as code completion ~\cite{Pavlichenko2025, SemenkinFLCC}. 
Yet, its alignment with developer flow is often overlooked at first ~\cite{RovdoKeynote,Barke2022}.
The LLMs powering AI features 
have increasingly better understanding of the task at hand through scaled training
~\cite{McKenna2025}, 
cleverer context-collection strategies
~\cite{SemenkinContext, Ustalov2025}, 
and advancements in fine-tuning processes. 
This leads to increasingly autonomous and proactive uses such as 
next-edit suggestions
~\cite{SemenkinNextEdit}
and agents
~\cite{Zakonov2025, Kovrigin2025}.
However, this focus on extending LLM-based functionality has inadvertently overshadowed a vital aspect of the user experience: the interaction dynamics between developers and their AI tools
~\cite{Sergeyuk2025}.

At the same time, macro-scale repository studies indicate rising duplication and code churn correlated with AI-assisted workflows
~\cite{Harding2025},
highlighting potential maintainability risks that are orthogonal to raw output volume. Evaluations of experienced developers show misalignment between perceived (+20\%) and actual productivity
(-20\%) ~\cite{Becker2025}.
A large-scale survey primarily finds AI in software development to be an ``amplifier'', both of software delivery throughput (+5\%) and of software delivery instability (+10\%)
~\cite{GoogleCloud2025}, 
underscoring the need to carefully calibrate AI-assisted tools to where they can be useful over the long term.


\subsection{Inline Code Completion}
We limit the scope of this paper to LLM-driven code completion in JetBrains IDEs. 
Our IDEs support both local-inferenced completions which work fully offline, through an on-device small LM designed for the in-editor language \cite{SemenkinFLCC}; and cloud-inferenced completions through larger LMs that can handle groups of semantically and syntactically similar languages \cite{Pavlichenko2025}. 
While this study is likely relevant to any similar tool, our focus is on the \textit{cloud-based code completion provided by JetBrains}. 

Our cloud-based code completion has about a ~30\% accept-per-show rate, which is similar to GitHub Copilot's 27.5\% in 2022 \cite{Ziegler2022}, indicating we are likely able to reduce two-thirds of generated completions without affecting the user experience drastically. Moreover, these shown completions constitute only one-third of completions that still match the code context (characters before the cursor) by the time they are generated. This implies we are \textit{theoretically} able to save two-thirds of inference requests without affecting the user experience at all, 
and close to 90\% of completions without affecting the number of accepted completions. 

There exist several studies investigating the useful and counter-productive interactions between developers and inline completion
\cite{Barke2022,Ziegler2022,MozannarReading,Prather2023,Vaithilingam2022}. To summarise, \citet{Barke2022} find that useful developer interactions are mainly bimodal:
either \textit{accelerative} where the developer knows what they want and uses the completions to get there faster, 
or \textit{explorative}, where the developer relies on the tool to suggest possible approaches \cite{deMoor2024}.
\citet{Prather2023} observe two additional modes primarily prevalent among novices: 
\textit{shepherding}, where a novice slowly accepts a series of suggestions, and 
\textit{drifting}, where the suggestions do not fully capture the user intent and they are led down a cyclic `debugging rabbit-hole'
~\cite{deMoor2024}. 

Designers of completion tools should further take into account the impacts on user flow: switching between acceleration and exploration modes adds cognitive overhead ~\cite{Barke2022}, 
often resulting in developers accepting suggested snippets while deferring their validation to a later point
~\cite{Ziegler2022}. 
The frequency and timing of these suggestions are critical to the overall productivity which the tool aims to boost 
~\cite{Prather2023}.
While some studies advocate for allowing users to configure the timing and context of completions to address these concerns
~\cite{Barke2022,Wang2023},
we argue that the vast majority of users likely expect these tools to work out-of-the-box and adapt to their usage patterns.

\subsection{Control Models for Code Completion}
Several earlier works propose ML classifiers that filter out unused or distracting completions to developers. 
A reverse-engineering blog-post reveals that GitHub Copilot has a logistic-regression trigger model 
\cite{Thakkar2022}.
\citet{MozannarWhen} (affiliated with GitHub) report that, together with a filter model, they are able to avoid generating 13\% of completions and filter 25\% of generated completions, out of which 95\% would have been rejected by the user. 
To this end, they use a variety of in-app telemetry features, and highlight the importance of personalised models aware of user mental state.

\citet{Sun2025} experiment with a transformer-based trigger model which can reject 20\% of completion requests, out of which 97.4\% would have resulted in unhelpful completions.

\citet{deMoor2024} show that both telemetry and text are useful information for a classifier. They combine in-app telemetry with code-context by processing it in a transformer model's output projection layer, reporting a 34\% filter rate with a 3\% false negative rate.

\section{Approach}
As shown in \Cref{fig:control-models-flow}, we aim to train two models: 
\textit{trigger} which controls when to inference a completion, and 
\textit{filter} which can leverage the completion features to potentially hide bad suggestions.
We consider a given completion to be positive if it is accepted by the user, and a negative if it is ignored or explicitly rejected through e.g. pressing the escape key.


\subsection{Model Architectures}

\subsubsection{Gradient-Boosting Models for Tabular Data}
We use the boosting implementation provided by CatBoost~\cite{Dorogush2018}, as it robustly handles categorical features. 
Boosting models are relatively fast at inference, which is ideal for a frequent interaction like code completion.

\subsubsection{Transformer-based Models for Code Context}
We further consider transformer classifiers to continue the line of research proposed by earlier work \cite{deMoor2024,Sun2025}. 
While their latency is relatively higher, transformers have a deeper contextual understanding of the scope around the cursor~\cite{deMoor2024}.

We build on the hybrid transformer design proposed by \citet{deMoor2024}, where the embedding of the classification token is concatenated with scalar features before being passed to the classification head. In our implementation, tabular features are first preprocessed following the approach described by \citet{Holzmller2024} and then encoded through a small multi-layer perceptron (MLP) prior to concatenation. Unlike their bidirectional encoder setup, we employ an in-house 100M-parameter code generation model as the backbone, which provides stronger out-of-the-box code understanding and alignment with our downstream task.

%
%

%


\subsection{Data Sources}
Our training and evaluation data consist of anonymised feature usage logs from 
Mellum~\cite{Pavlichenko2025} cloud-inferenced completion.

\begin{table}
  \caption{
    Evaluation Datasets.
    Label imbalance is given as the number of negative labels per positive label.
    ``Context'' is a separate Kotlin dataset with raw code context.
  }
  \label{table:evaluation-datasets}
  \begin{tabular}{lrrrr|r}
    \toprule
                    & Kotlin  & Python  & PHP   & C\#   & Context   \\
    \midrule
    $n$ users       & 4.2k    & 11.9k   & 12.5k & 9.7k  & 98        \\
    generations     & 61k     & 122k    & 178k  & 125k  & 28k       \\
    \textit{label imbalance}                                        \\
    \;\; trigger    & 11.5    & 9.1     & 8.7   & 10.5  & 15.0      \\
    \;\; filter     & 2.5     & 2.4     & 2.2   & 2.5   & 2.3       \\
    \bottomrule
  \end{tabular}
\end{table}

\subsubsection{Tabular Telemetry Features}
\label{sec:tabular-telemetry-features}

Tabular telemetry logs consist of categorical and scalar features for every completion event.
Each sample consists of several hundred features. 120 features are always collected at the start of the event, and they can be leveraged by the trigger model, e.g. the user's current typing speed.
Samples that are relevant for the filter model have a few hundred more features about the collected context, model execution, and the completion itself.
To cover a range of differing syntax and semantical density, we consider four languages:
Kotlin, Python, PHP, and C\#. 

Some statistics about the evaluation dataset are given in \Cref{table:evaluation-datasets}.
Due to the staggered nature of real-world software development, we are able to ensure a consistent dataset across our models only for their evaluation. 
This is because, to train new models, we need to disable the existing ones to avoid a biased label distribution.

\subsubsection{Code Context}
\label{sec:code-context-features}

For our transformer-based models, we collect raw code context alongside tabular telemetry features; 
see \Cref{table:evaluation-datasets}.
At the time of writing, we only collect Kotlin code context for internal users, and we cannot evaluate models trained on this dataset online, 
due to the risks of leaking sensitive information.
We discuss this limitation and workarounds in \Cref{sec:obstacles-for-local-llms}.

However, for this dataset, we managed to ensure a consistent train and test split across all our models, 
including the boosting-based models trained solely on tabular telemetry features.
Furthermore, we can guarantee that the set of users in the test split is disjoint from those in the train split. 
We have 126 users in the train split, totalling about 70k generations; 
and 98 users in the test split, totalling around 28k generations 
(\Cref{table:evaluation-datasets}; right-most column).



\subsection{Metrics}
\label{sec:metrics}

We discuss our metrics working backwards from what we can measure in the online user experience, 
towards proxies for these metrics that can be used in offline evaluation. 

\begin{itemize}
  \item \textit{Ratio of Completed Code} (RoCC)
    is our main metric for assessing code completion quality \cite{SemenkinFLCC}. 
    It is defined as the ratio between the number of symbols inserted by code completion and the total number of symbols written. 

  \item \textit{Accept Rate} (AR)
    is the number of accepted completions divided by the number of shown completions. 
    A completion is accepted when the user presses the \texttt{tab} key.
    \citet{Ziegler2022} show that the acceptance rate strongly correlates with perceived productivity.

  \item \textit{Cancel Rate} (CR)
    is the number of cancelled completions divided by the number of shown completions. 
    Note that we do not consider all ignored completions, e.g. when the developer continues typing. 
    Rather, we track \textit{explicit} cancellations, which consist of pressing the escape key, mouse clicks, or caret movement;
    as they are more likely to indicate user frustration due to e.g. distracting completions.

\end{itemize}

In the offline setting, we modify \textit{Ratio of Completed Code} to be only its numerator: the number of \textit{Symbols Completed} by code-completion.
This is because of the tight dependence between inline completion events, 
let alone how they influence other actions in the editor; 
i.e. these events are not independent nor identically distributed.
This limitation is discussed in \Cref{sec:discussion-dependence-of-completion-events}.


\section{Setup \& Results}
We first evaluate our control models offline, and only move to the online setting when results are promising enough and deployment satisfies legal requirements.
This reflects the natural development process from proof-of-concept to a production environment, where the models may affect millions of users.
Our contributions can be summarised into three main research questions as follows:

\begin{itemize}

  \item[RQ 1] 
    How does the classification performance of boosting-based control models stack up against a transformer-based approach?
    Using our \textit{Code Context} dataset (\Cref{sec:code-context-features})
    we investigate how many generations can be filtered at different false-negative rates 
    ($\text{FNR} \in \{0.01, 0.05, 0.10, 0.20\}$)
    for each model architecture
    and the impact on our completion quality metrics.


  \item[RQ 2] 
    How do boosting-based control models perform \textit{offline} on a set of popular, but syntactically diverse languages? 
    We use our \textit{Tabular Telemetry Features} dataset (\Cref{sec:tabular-telemetry-features}), 
    for Kotlin, Python, PHP, and~C\#.
    Similar to RQ1,
    we pin the false-negative rate FNR= 0.10, and look at how many
    inference requests the trigger model can prevent, as well
    as the estimated impact on metrics after completions pass through
    the filter model.


  \item[RQ 3] 
    How do boosting-based control model metrics translate to the \textit{online} environment, 
    where the independence between completion sessions no longer holds? 
    We report results from our latest A/B studies on languages for which it was performed: Java, Python, and Kotlin for Filter Models, and Kotlin for Trigger Models.
    While this is a smaller dataset than the previous RQ, 
    we hope to illustrate which findings can be generalised from our offline evaluation to the online environment.

\end{itemize}

\subsection{Offline Classification Performance of Boosting and Transformer Control Models}

\begin{figure*}[]{}
  \label{fig:boosting-vs-transformers}
  \centering

  \begin{subfigure}[b]{0.33\textwidth}
    \includegraphics[width=\textwidth]{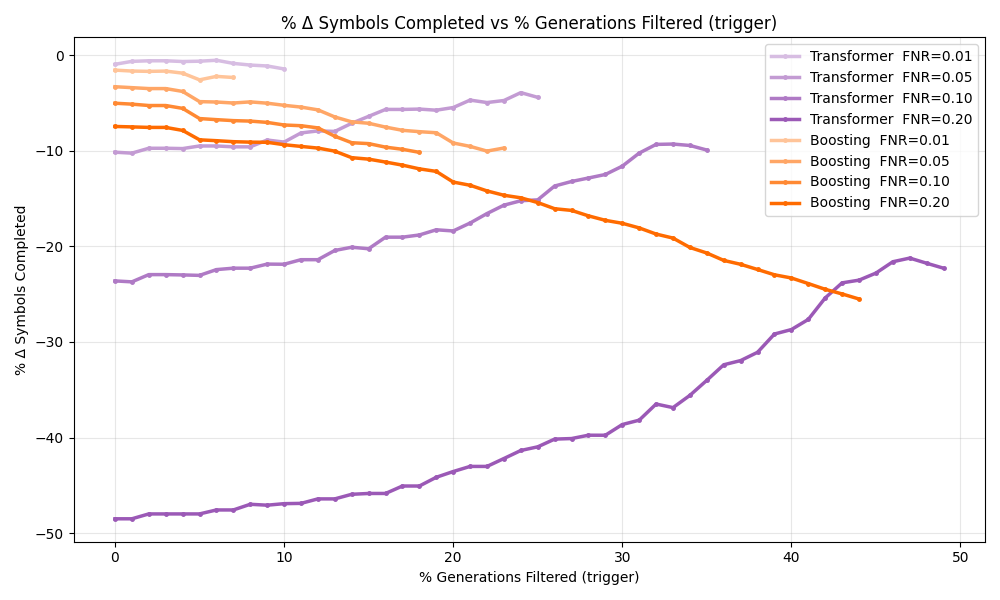}
    \caption{Symbols Completed}
    \label{fig:boosting-vs-transformers-symbols-completed}
    \Description[
      Symbols Completed for Boosting and Transformer Control Models, 
      at different false-negative rates, and percent of generations filtered.
    ]{
      Symbols completed decreases steadily as more control is given to the boosting trigger model, for all FNRs. 
      However, the same does not hold true for the transformer-based approach: 
      as more control is given to the trigger model, the amount of completed symbols increases.
    }
  \end{subfigure}
  \hfill
  \begin{subfigure}[b]{0.33\textwidth}
    \includegraphics[width=\textwidth]{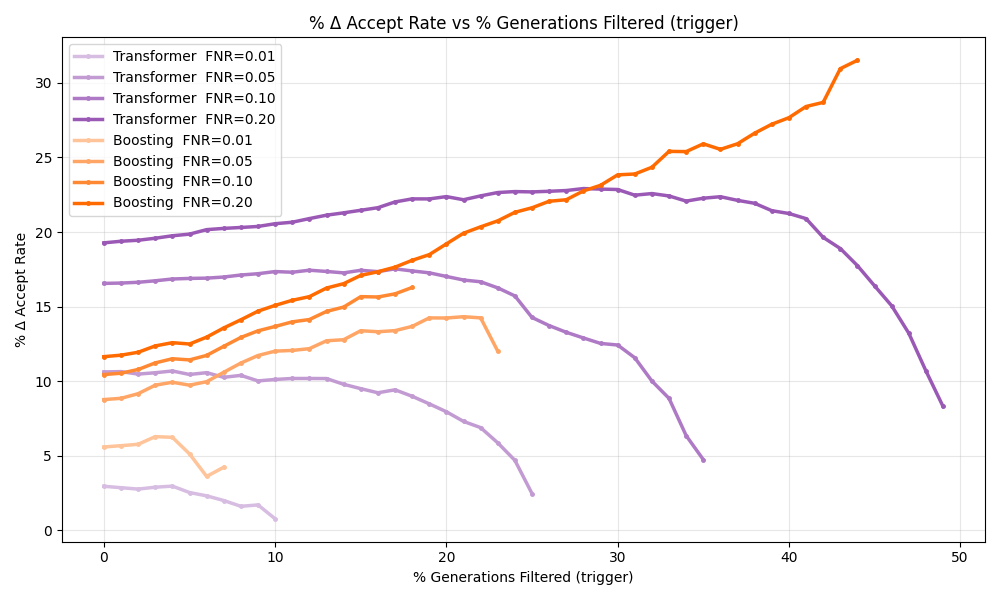}
    \caption{Accept Rate}
    \label{fig:boosting-vs-transformers-accept-rate}
    \Description[
      Accept Rates for Boosting and Transformer Control Models, 
      at different false-negative rates, and percent of generations filtered.
    ]{ 
      Accept Rates increase for Boosting Approach as more generations are reduced by the trigger model. 
      For the Transformer Approach, they increase up to a point, 
      after which it seems more desirable to leave them to the filter model
    }
  \end{subfigure}
  \hfill
  \begin{subfigure}[b]{0.33\textwidth}
    \includegraphics[width=\textwidth]{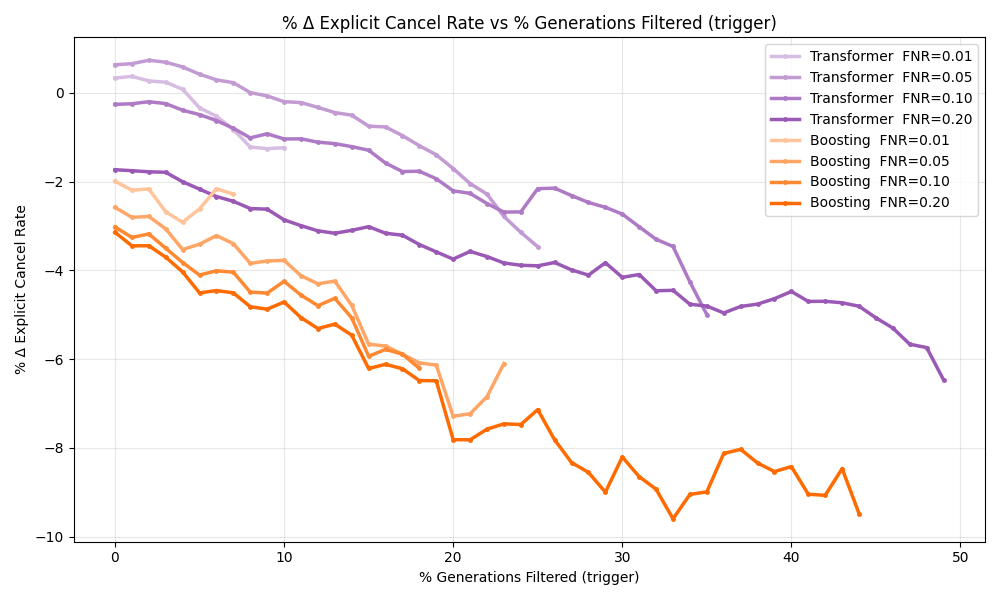}
    \caption{Cancel Rate}
    \label{fig:boosting-vs-transformers-cancel-rate}
    \Description[
      Cancel Rates for Boosting and Transformer Control Models, 
      at different false-negative rates, and percent of generations filtered
    ]{
      Cancel Rates decrease for Boosting Approach as more generations are filtered by the trigger model. 
      Cancel Rates similarly decrease for the Transformer approach, albeit at a lower rate as the percent of generations filtered by the trigger model increases.
    }
  \end{subfigure}

  \caption{
    Impact of Boosting- and Transformer-based Control Models 
    on Completion Quality Metrics, 
    at Different Generation Filter Rates and Overall False-Negative Rates (FNR), 
    Computed Offline.
    How to Read These Graphs: More control is handed over to the trigger model as \textit{\% Generations Filtered} increases.
  }
\end{figure*}

We compare the classification performance of boosting and transformer control models, 
and investigate the interplay between our trigger and filter models.
As seen in \Cref{fig:boosting-vs-transformers-symbols-completed}, 
there are points where transformer-based control models are able to filter out more generations while taking less of a $\Delta$ \textit{Symbols Completed} hit than the boosting-based approach.

However, transformer-based models, at these higher filter rates, are not able to improve the Accept and Cancel Rates 
(\Cref{fig:boosting-vs-transformers-accept-rate} and \Cref{fig:boosting-vs-transformers-cancel-rate})
to the same extent as boosting models.
This is because the denominator of these rate-based metrics (shown completions) is relatively less affected than for the boosting-based approach. 
While we could report the impact on the absolute number of accepts and cancels, 
this is intentional, as it models the perceived effect to the end users of our completion service.
We invite the reader to think about how to balance the opposing objectives of improving Accept rate and Cancel rate metrics while not harming the number of Symbols completed.

\subsection{Offline Classification Performance of Boosting Models on Different Languages}

\begin{figure*}[]{}
  \label{fig:boosting-on-different-languages}
  \centering

  \begin{subfigure}[b]{0.33\textwidth}
    \includegraphics[width=\textwidth]{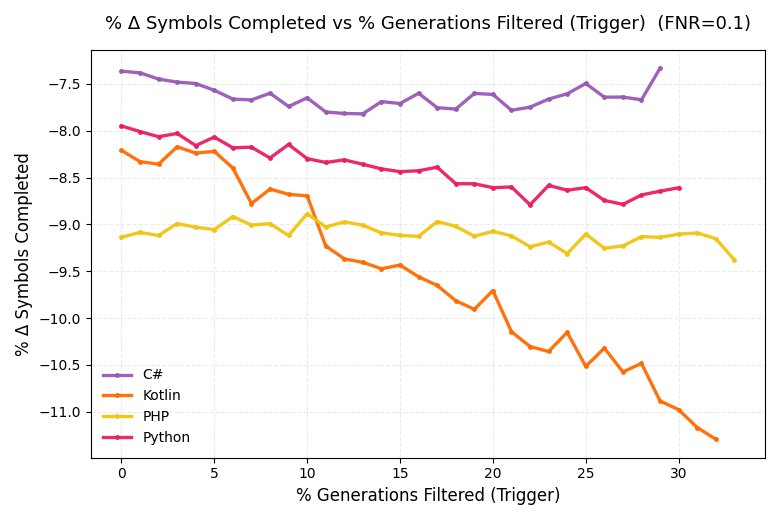}
    \caption{Symbols Completed}
    \label{fig:boosting-on-different-languages-symbols-completed}
    \Description[
      Symbols Completed for Boosting Control Models, 
      at 0.10 false-negative rate, and percent of generations filtered;
      for a range of languages.
    ]{
      Symbols Completed decreases as we hand control over to the trigger model. 
    }
  \end{subfigure}
  \hfill
  \begin{subfigure}[b]{0.33\textwidth}
    \includegraphics[width=\textwidth]{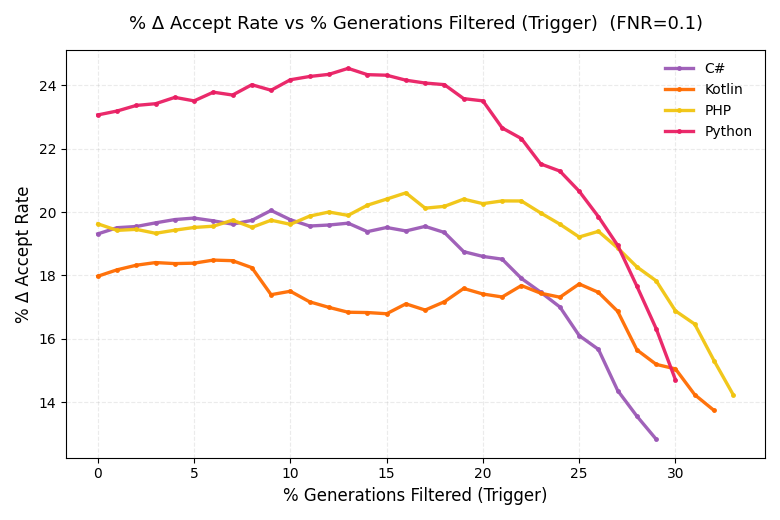}
    \caption{Accept Rate}
    \label{fig:boosting-on-different-languages-accept-rate}
    \Description[
      Accept Rates for Boosting Control Models, 
      at 0.10 false-negative rate, and percent of generations filtered;
      for a range of languages.
    ]{ 
      Accept rates remain roughly constant until about 20\% generations filter rate, 
      after this, it is not wise to hand more control over to the trigger model.
    }
  \end{subfigure}
  \hfill
  \begin{subfigure}[b]{0.33\textwidth}
    \includegraphics[width=\textwidth]{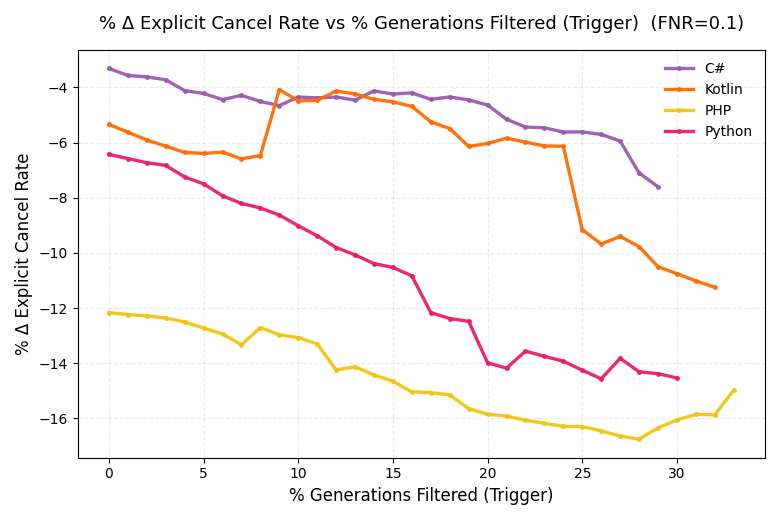}
    \caption{Cancel Rate}
    \label{fig:boosting-on-different-languages-cancel-rate}
    \Description[
      Cancel Rates for Boosting Control Models, 
      at 0.10 false-negative rate, and percent of generations filtered;
      for a range of languages.
    ]{
      Cancel Rates decrease as we hand more control to the trigger model
    }
  \end{subfigure}

  \caption{
    Impact of Boosting-based Control Models 
    on Completion Quality Metrics, 
    at Different Generation Filter Rates and Overall 0.10 False-Negative Rate (FNR), 
    for Different Languages, Computed Offline.
    How to Read These Graphs: More control is handed over to the trigger model as \textit{\% Generations Filtered} increases.
  }
\end{figure*}

As user interactions with code-completion differ per language \cite{Thakkar2022}, we investigate the effect of our control models on different languages and observe some interesting differences. 
For instance, \Cref{fig:boosting-on-different-languages-symbols-completed} indicates that for Kotlin, it makes sense to hand more control over to the filter model.
We also note that the perceived impact on 
Accept Rate (\Cref{fig:boosting-on-different-languages-accept-rate}) 
and Cancel Rate (\Cref{fig:boosting-on-different-languages-cancel-rate})
differs considerably from language to language.

Intuitively, the filter model should be a better classifier than the trigger model, 
as it has access to more information, namely the features of the generated completion. 
Curiously however, there are points for which it makes more sense to hand over control to the trigger model instead of the filter model when sacrificing a given number of false-negatives, 
e.g. for PHP in \Cref{fig:boosting-on-different-languages-symbols-completed}.
This could be because we are limited to a much smaller training set for the filter model, 
as disabling these models in production results in a serious adverse impact on the user experience.

There also exists an abnormality in Kotlin's Cancel Rate in \Cref{fig:boosting-on-different-languages-cancel-rate}.
Our main hypothesis for it is that as the trigger model's threshold moves in discrete steps corresponding to the percentiles on the x axis. 
Each step can suddenly change the mix of examples, especially if positives are not evenly distributed.

%

\subsection{Online Classification Performance of Control Models}
\label{sec:online-classification-performance}

We ablate our boosting-based trigger and filter models in a production environment using A/B tests, 
and report user-level metrics in \Cref{table:online-evaluation}.
The metrics marked with an asterisk (*) are not statistically significant at $p \le 0.05$, 
where significance is computed using bootstrap resampling with individual users as the sampling unit.
For the filter model, we also apply a few hard rules as explained by \citet{SemenkinFLCC}, 
such as filtering for non-compilable code. 
Even though the output of such static analyses is an input feature to the filter model, 
this likely inflates the metrics slightly compared to the offline evaluation.

Filter models improve Accept and Cancel Rates, likely enhancing the user's perception of the completion, 
as the frequency of accepts is what primarily drives user's perceived productivity \cite{Ziegler2022}.
However, it also comes at the cost of reducing the \textit{Ratio of Completed Code}; 
corroborating our offline findings. 

For the trigger model ablation, both A/B groups contain a filter model. 
As removing it would likely significantly affect the user experience, we instead perform A/B on a larger user group with the filter model present.
Curiously, while it similarly affects completion quality metrics on a smaller scale than the filter model (as expected), 
we note that it did not significantly reduce the number of generations. 
This is despite it preventing around 20\% of completions from being generated. 
We further discuss this in \Cref{sec:discussion-dependence-of-completion-events}.

\begin{table}[H]
  \caption{
    Relative Change ($\% \Delta$) in Online A/B Evaluation Metrics 
    for Boosting-based Trigger and Filter Models.
  }
  \label{table:online-evaluation}
  \begin{tabular}{lrrrrrr}
    \toprule
    Model                   &         & \multicolumn{3}{c}{Filter} & & Trigger    \\
    Language                &         & Java      & Python    & Kotlin  & & Kotlin     \\
    \midrule

    $n$ users (A/B)         &         & 151/127   & 111/94    & 95/62   & & 1740/1771  \\
    \midrule

    $\% \Delta$ Generations &         & n/a       & n/a       & n/a     & & -13.8    \\

    $\% \Delta$ RoCC        &         & -9.6*   & -14.0*  & -13.9*   & & -2.4       \\

    $\% \Delta$ AR          &         & +46.5    & +32.9    & +47.6      & & +2.7       \\

    $\% \Delta$ CR          &         & -36.7   & -15.5*  & -35.0*      & & -4.5 \\


    \bottomrule

  \end{tabular}
\end{table}

\section{Discussion on Metrics}

\subsection{The Dependence of Completion Events}
\label{sec:discussion-dependence-of-completion-events}
As mentioned in \Cref{sec:metrics}, metrics computed during offline evaluations do not necessarily translate to the online setting, 
due to the strong dependencies between sequential code completions in a given programming session. 
While we have considered it out-of-scope, 
future work could consider modelling these dependencies by investigating correlations between events.

For instance, despite selecting a threshold that filters around 20\% of the completions reaching our online trigger model (\Cref{sec:online-classification-performance}), 
we only reduced the amount of generated completions per-user by 13.8\%. 
The dependence between completion events is evident here: 
\textit{by affecting completion behaviour, users change their interactions with code completions}, 
resulting in more overall opportunities for completions. 
This prompts us to evolve the metrics we track. 



\subsection{Measuring Long-Term Productivity Gains}
\label{discussion-of-online-metrics}

Related work finds that our current choice of metrics primarily aims to improve the \textit{perceived} productivity of developers
~\cite{Ziegler2022,Barke2022}, 
while other studies raise questions about whether this perception actually materialises in practice 
~\cite{Becker2025,GoogleCloud2025,Harding2025}.
There is certainly a need to track longer-term impacts of in-IDE AI assistance, but this begs the question:
what is an appropriate time-frame to consider for such metrics?

We prompt the reader to consider this. 
On the one hand, optimising with respect to shorter time-frames likely improve developers' perception and enjoyment of features such as code completion, 
as indicated by the effective `accelerative' behaviour observed by \citet{Barke2022}.
On the other hand, it is unlikely that users are willing to pay for a service that makes developers think they are 20\% faster, when in effect it takes them 20\% longer to complete a given task with AI assistance \cite{Becker2025}.

%

\section{Future Work}

%
%

\subsection{Obstacles for Transformer-based Approach}
\label{sec:obstacles-for-local-llms}
The primary obstacle to adapting transformer-based control models is that it is hard to guarantee latency constraints on our users' vast range of consumer hardware.
Furthermore, integrating code context alongside tabular-like features requires non-trivial modifications to the inference backend 
such as \texttt{vLLM} (for cloud-based inference) and \texttt{llama.cpp} (for local inference). 
For these reasons, there is a clear trend in adoption of smaller and faster CatBoost-based control models inside JetBrains IDEs \cite{Dorogush2018}.

Transformer-based models also present security and privacy risks. 
Such models can memorise and later reveal sensitive information from their training data, even when adapted solely for classification purposes
~\cite{Elmahdy2022}.
This could be addressed by applying privacy-aware training techniques such as differential privacy
~\cite{McKenna2025}.

\subsection{Modelling \& Optimisation Directions}

CursorTab reports a promising reinforcement-learning setup that continuously improves suggestion policies from live accept/reject feedback~\cite{Jackson2025}. This line of work indicates that gating can be learned end-to-end from interaction signals, potentially removing the need for a separate trigger model.
It could be beneficial to explore a single model that generalises across different use cases (e.g. trigger and filter), which would simplify training, serving, and maintenance.

\subsection{Designing for User Personalisation}
At the time of writing, we see the ``lowest-hanging fruit'' for architectural improvements to be designing for user personalisation, corroborating similar works \cite{XuChen2019,MozannarWhen}. 
Manual inspection of wrongly classified samples showed that selections and rejections are often a matter of personal preference. 
JetBrains IDEs collect a variety of time-series telemetry, outside the code-completion interaction, that could be used for modelling such user preferences. 

There exists some work that successfully predicts the next user state in JetBrains IDEs \cite{Koohestani2025}.
Further work may be able to draw inspiration from preference modelling research used in domains such as e-commerce \cite{Zhou2024}.
By underscoring this research direction, we hope to inspire work in this area.


\begin{acks}
We're particularly thankful to 
Ivan Dolgov, Anton Semenkin, and Uladzislau Sazanovich for motivating this direction;
Sergey Titov and Evgeny Grigorenko for seeing the potential in trigger models;
Gleb Marin and Vladimir Fedorov for their open-source development of our ML API; 
as well as Daniil Bubnov, Kirill Karnaukhov, and Kirill Krylov for their integration support in our IDEs.
\end{acks}

\bibliographystyle{ACM-Reference-Format}
\bibliography{export}

\appendix

\end{document}